# UNNATURAL-PARITY $(\vec{p},\vec{p}')$ REACTIONS IN A FACTORIZED IMPULSE-APPROXIMATION MODEL FOR POLARIZATION TRANSFER (PT) AND SPIN RESPONSES[*]

Plavko A.V.[1], Onegin M.S.[2], Kudriashov V.I.[3]
[1] *St. Petersburg State Polytechnic University, Russia;*
[2] *Petersburg Nuclear Physics Institute, Gatchina, Russia;*
[3] *St. Petersburg State University, Russia*

Linear combinations, $D_K$, of the complete polarization-transfer observables, $D_{ij}$, in the $(\vec{p},\vec{p}')$ reaction at intermediate energies are demonstrated. A comparison between systematized measured and calculated values of the combinations $D_K$ for the $1^+$ ($T = 0$ and $T = 1$) levels in $^{12}$C, for the $4^-$ ($T = 0$ and $T = 1$) states in $^{16}$O, and for the $6^-$ ($T = 0$ and $T = 1$) levels in $^{28}$Si are reported. Particularities in angular distributions of transverse- and longitudinal-spin-transfer probabilities, $D_K$, for the $T = 0$ and $T = 1$ unnatural-parity states in the indicated nuclei are discussed. The spin-observable combinations $D_{ls}$ allowed to differentiate reliably the strength of the isoscalar and isovector spin-orbit interactions. The comparison of experimental $D_K$ and calculated $D_K$ with the use of zero-range treatment (LEA code) and exact finite-range calculations (DWBA 91 program) made it possible to identify the role of exchange contributions.

## INTRODUCTION

The purpose of this paper is to continue the parametrization of complete spin-transfer measurements for inelastic polarized-proton scattering on the basis of the polarization observables $D_K$ introduced by Bleszynski et al. [1] and expressed in terms of the conventional Wolfenstein parameters [2].

Complete measurements of the spin-transfer coefficients $D_{ij}$ require beams of protons with initial polarization, longitudinal ($L$), normal to the reaction plane ($N$), and sideways ($S = N \Box L$). The spin (polarization)-transfer observable $D_{ij}$ relates to the component $i$ of the incident (initial) proton polarization, and to the component $j$ of the outgoing (final) proton polarization [3, 4]. The components $D_{ij}$ (alternate parameters suggested by Wolfenstein [2]) are the experimental measurements.

In theory, according to [1, 2], the spin-transfer coefficients $D_{ij}$ can be defined in terms of the $NA$ (nucleon–nucleus) scattering amplitude $\bar{M}(q)$ and nucleon spins:

$$D_{ij} = Tr\,[\bar{M}\sigma_i \bar{M}^+\sigma_j]/2I_0. \qquad (1)$$

Here $\sigma_i$ and $\sigma_j$ are the Pauli spin matrix with respect to the $i$ ($L$, $N$, $S$) and $j$ ($L'$, $N'$, $S'$) axes. The prime on the second (outgoing) subscript is normally omitted. $I_0$ is the usual unpolarized differential cross section.

The $NA$ scattering amplitude $\bar{M}(q)$ or, in other words, the collision operator for the $(\vec{p}, \vec{p}')$ transition in PWIA can be expressed in terms of the free $NN$ scattering amplitude operator in the nonrelativistic framework $M(q)$ [1–6].

The use of polarization-transfer observables $D_{ij}$ in (1), and especially that of the combinations of these observables $D_K$, obtained, according to Bleszynski et al [1], in inelastic proton scattering allows to employ this scattering (see [7]) as a filter to examine particular pieces of the effective $NN$ scattering.

Model calculations for $\bar{M}(q)$ are represented in detail specifically in [5]. It is worth noting that in the case of spin-flip transitions at an angle of 0°, the $NN$ scattering amplitude in the PWIA approximation becomes significantly simplified [6].

The motivation for the authors of paper [1] was that they believed that the Wolfenstein parameters [2] were not related in a transparent way to the $(\vec{p}, \vec{p}')$ collision matrix. They introduced a particular set of observables $D_K$ that were expressed in terms of linear combinations of the $D_{ij}$ parameters. Each of these observables $D_K$ (where $K = ls, q, n, p$), according to the authors of work [1], depended separately on the specific tensor components of the $(\vec{p}, \vec{p}')$ collision matrix. Following these model representations, a different parametrization should be also applied to experimental results, alternative to the conventional Wolfenstein parameters.

Consequently, the $D_K$ observables can be defined in terms of components of the general collision matrix and also in those of experimental $D_{ij}$ parameters. The authors of [1] managed to obtain, in a single collision approximation, expressions for the observables $D_K$ that displayed dependence of an individual $D_K$ on particular components of the $NN$ amplitude and on particular nuclear form factors (see [3–11]).

Such an approach to the case of nucleon-nucleus scattering was highly appraised in [3–11] since it demonstrated that spin observables could be used to isolate specific parts of the nuclear response. In [3–11] especially highly rated was the fact that that approach opened up the possibility to separate the transverse- and longitudinal-spin-response functions of the nuclei by measuring spin (polarization)-transfer observables with the use of polarized beams.

# 1. PARAMETRIZATION OF COMPLETE SPIN-TRANSFER MEASUREMENTS IN $(\vec{p},\vec{p}')$ SCATTERING FOR UNNATURAL-PARITY STATES IN $^{12}$C, $^{16}$O, AND $^{28}$Si NUCLEI

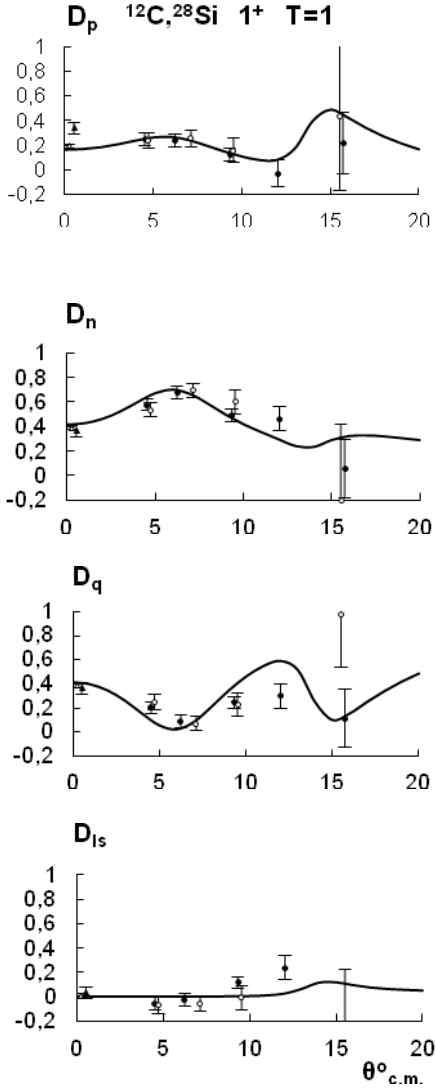

As it was noted above, four polarization observables $D_K$, introduced in [1], can be expressed in terms of linear combinations of the Wolfenstein parameters [2], or analogous complete PT coefficients ($D_{ij}$) for the $(\vec{p},\vec{p}')$ reaction. In PWIA with optimal factorization approximation, the combinations $D_K$ are given as:

$$D_{ls} = [1 + D_{NN} + (D_{SS} + D_{NN})\cos\theta - \delta\sin\theta]/4,$$
$$D_q = [1 - D_{NN} + D_{SS} - D_{LL}]/4,$$
$$D_n = [1 + D_{NN} - (D_{SS} + D_{LL})]\cos\theta + \delta\sin\theta/4, \qquad (2)$$
$$D_p = [1 - D_{NN} - D_{SS} + D_{LL}]/4,$$

where $\theta \equiv \theta_{c.m.}$ (deg.) is a scattering angle, and $\delta = D_{LS} - D_{SL}$.

*Fig. 1. Systematized angular distributions of the spin-transfer probabilities, $D_K$, for the isovector $1^+$ (15.11 MeV) level in $^{12}$C, and $1^+$ (11.5 MeV) level in $^{28}$Si based on the measured $(\vec{p},\vec{p}')$ quantities at 500 MeV (filled circles [3]), open circles [4]), and at ~ 400 MeV [6] (open triangles – $^{12}$C, filled triangles – $^{28}$Si). Our calculations (curves) are described in the text.*

$D_{ls}$ is associated with the spin-orbit term in the *NN* effective interaction. The other three $D_K$ are associated with the tensor terms for each axis as $D_q$ with the momentum transfer $\vec{q} = \vec{k}_{in} - \vec{k}_{out}$, $D_n$ – with normal $\vec{n}$ to the reaction plane, and $D_p$ – with $\hat{p} = \hat{n} \times \hat{q}$.

As an example, Fig. 1 shows experimental and calculated dependences for all four $D_K$ combinations for the $1^+$, $T = 1$ level in $^{12}$C and $^{28}$Si.

The novelty of the present work is that the range of the measured spin-observable combinations, $D_K$, has been extended to 0°, which allowed for the first time to evaluate the validity of the analytical results at extremely forward angles (at and near zero degrees). Besides, the measured $D_K$ for the $1^+$, $T = 1$ state in $^{12}$C at 0° have been completed by similar data for the $1^+$, $T = 1$ level in $^{28}$Si (Fig. 1).

The above-mentioned combinations, $D_K$, become simpler at small scattering angles since $\delta \approx 0$, $\cos \theta \approx 1$ and $\sin \theta \approx 0$. According to [6], in the case of 0° for the isovector ($\Delta T = 1$) $M1$ transition we have $D_{SS} = D_{NN} \approx D_{LL} \approx -1/3$. If these quantities are inserted into the simplified expressions $D_K$, we easily get the following set: $D_{ls} = 0$, and $D_q = D_n \approx D_p \approx 0.3$. Furthermore, these quantities appear to be approximately equal to the measured values and to our DWIA calculations using the FL (Franey and Love) interaction and the program LEA from Kelly. This is quite similar to what is shown in Fig. 1.

As for the description of the $D_K$ data at other angles, the program LEA, used here for the first time, turned out to be more effective overall than some other programs proceeded from nonrelativistic and relativistic calculations that were previously employed for such purposes [3].

As is seen in Fig. 1 for $T = 1$, the angular distributions of the transverse- and longitudinal spin-transfer probabilities, $D_n$ and $D_q$, respectively, are considerably different in shape, as they are in the case of a lower energy ($E_p = 200$ MeV), which we will discuss further on. This phenomenon can be explained by a significant difference in the momentum dependence of transverse- and longitudinal axial form factors.

At the same time the angular distributions, $D_n$ and $D_q$, in calculations and experiments, have similar smooth shapes both at $E_p = 400$–500 MeV and at 200 MeV for the isoscalar $1^+$ (12.71 MeV) state in the $^{12}$C$(\vec{p}, \vec{p}')^{12}$C reaction [8]. This is in good agreement with the fact that the moduli of the isoscalar interaction components of the free nucleon–nucleon $t$-matrix for 140-Mev (and 800-Mev) nucleons from Love and Franey, responsible for transverse and longitudinal transitions, respectively, are similar in value and are rather flat as functions of the quantity $q$ [5].

## 2. PARTICULAR COMBINATIONS $D_K$ AND SPECIFIC AMPLITUDES OF THE EFFECTIVE INTERACTION

According to the model of Bleszynski et al. [1], certain combinations of $D_{ij}$ should demonstrate particular selectivity to the specific components of the $(\vec{p}, \vec{p}')$ collision matrix. Basically, it comes to the fact that the four functions $D_K$ (2) appear to be sensitive primarily to individual terms in the $NN$ interaction. In this case, the $NN$ interaction is normally the $t$ matrix in KMT (Kerman–MacManus–Thaler) representation [9].

The three polarization observables $D_K$ are predominantly sensitive to the response of the nucleus through the tensor operators that act along three axes: $\sigma_{1q}\sigma_{2q}$, $\sigma_{1n}\sigma_{2n}$, and $\sigma_{1p}\sigma_{2p}$. Here, as it was indicated above, $\hat{n}$ is normal to the scattering plane, $\hat{q}$ lies along the momentum transfer to the projectile, and $\hat{p} = \hat{n} \times \hat{q}$. The observable $D_{ls}$ is sensitive to the response through the spin-orbit operator ($\sigma_{1n} + \sigma_{2n}$). Then are used all four spin-dependent KMT amplitudes: $E$, $B$, $F$, and $C$ for $D_q$, $D_n$, $D_p$, and $D_{ls}$, respectively (see, e.g., [10]). The amplitudes $E$, etc., are complex and have isospin dependence, with the isospin index (0 or 1) being omitted.

According to [1] and [4], the particular combinations $D_K$ (2) in simplified assumptions can be related to specific amplitudes of the effective interaction and nuclear-structure-dependent components via

$$D_{ls} \approx |C|^2 [\chi_T]^2 / I_0 \qquad D_n \approx |B|^2 [\chi_T]^2 / I_0$$
$$D_q \approx |E|^2 [\chi_L]^2 / I_0 \qquad D_p \approx |F|^2 [\chi_T]^2 / I_0. \qquad (3)$$

Here $\chi_T$ and $\chi_L$ are transverse and longitudinal form factors, and $I_0$ is an unpolarized cross section.

We employed Equations (3) for the case of $T = 0$ in [8], and in this paper, we extended it to the excitation of states with $T = 1$. We also tested it at $\theta = 0°$, where $M(q)$ was so simplified that the two-body spin-orbit term could be neglected ($C = 0$), as well as the direct tensor interaction. In the case of no tensor interaction, we got the ratio $B = E$ (see, e.g., [11]).

In general, $\chi_T$ and $\chi_L$ are coherent sums of matrix elements for the transfers of different $\Delta L$. In the case of a dominant single $\Delta L$, the ratio $\chi_T^2/\chi_L^2$ can be simply replaced by a number [12]. In the present paper, we analyze such particular cases.

For $M1$ transitions ($\Delta J = 1$) at very forward angles, including 0°, the contribution with $\Delta L = \Delta J - 1$ dominates. For the excitation of the $1^+$, $T = 0$ and $1^+$, $T = 1$ states, dominates $\Delta L = 0$ (the $\Delta L = \Delta J + 1$ contribution is negligible) [6]. This ensures that in this case

$$\chi_T^2 = \chi_L^2. \qquad (4)$$

Taking all the above characteristics of Equation (3) into account, we get the following results at 0°:

$$D_{ls} \approx 0, \; D_q \approx D_n. \qquad (5)$$

It is this particular phenomenon that we observe in Fig. 1 both in the experiment and in our DWIA calculations for $T = 1$. We established Ratio (5) in [8] for the excitation of the $1^+$, $T = 0$ state in $^{12}$C. The only difference is that at the $\Delta T = 0$ transition the coefficients $D_q$ (and $D_n$) acquire different values, which is natural due to the dependence of all the members of *M(q)* on the isospin. In this way, Ratio (3) at 0° can be reliably tested and appear to be quite adequate.

## 3. ISOSPIN AND ANGULAR DEPENDENCIES OF SPIN-TRANSFER PROBABILITIES $D_{ls}$ IN UNNATURAL-PARITY $(\vec{p}, \vec{p}')$ REACTIONS AT INTERMEDIATE ENERGIES

In this section, we demonstrate that the observables $D_{ls}$, based on a set of polarization-transfer measurements, and their analysis within the framework of the model of Bleszinski et al. are applicable for a systematic evaluation of the role of spin-orbit interactions in nucleon inelastic scattering on a number of light-weight nuclei.

Accordingly, numerous research data, e.g., [3, 13, 14], prove that the isovector spin-orbit interaction is consistently weak at intermediate energies. On the other hand, the isoscalar spin-orbit component of the effective interaction is large. It is in agreement with the fact that the corresponding spin-transfer probabilities, $D_{ls}$, are primarily driven by the strengths of spin-orbit amplitudes (Fig. 2).

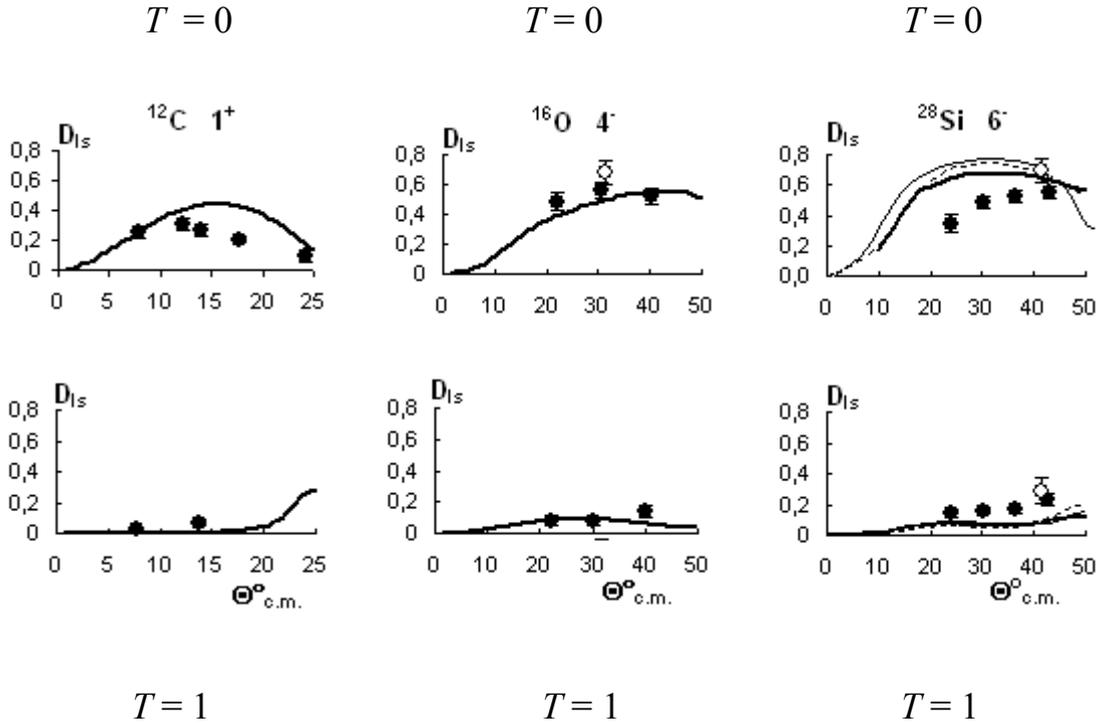

***Fig. 2.*** *Systematized experimental (points) and predicted (curves) data $D_{ls}$ for ($\vec{p}, \vec{p}'$) scattering on $^{12}C$ ($1^+$), $^{16}O$ ($4^-$) and $^{28}Si$ ($6^-$) to the T = 0 (top) and T = 1 (bottom) excitations. We based the represented experimental results $D_{ls}$ on ($\vec{p}, \vec{p}'$) spin-observable measurements at $E_p$ = 200 MeV [15, 16] (solid points) and at $E_p$ = 350 MeV [7] (open point – $^{16}O$), with the added data $D_{ls}$ at $E_p$ = 500 MeV [17] (open points – $^{28}Si$). Due to the difference in energies $E_p$, we introduced certain kinematic corrections. The calculated data (curves), described in the text, refer to $E_p$ = 200 MeV, apart from one incident for $^{28}Si$ (500 MeV).*

In the case of $^{12}$C we employed DWIA calculations (solid curves) with the DBHF interaction ([15, 16] and a private communication from Stephenson). For $^{16}$O, we performed calculations using the DWBA 91 program from Raynal with the $G$-matrix from Geramb (solid curve). For $^{28}$Si, all our calculations were made using the same DWBA 91 program with three types of interactions: the $G$-matrix from Geramb (solid curve) and two versions of the Idaho interaction [18] for $T = 1$ (almost undistinguishable in Fig. 2). In the case of $T = 0$ for $^{28}$Si we used one of the versions of the Idaho interaction and LEA FL at 500 MeV (thin solid and dashed curves, respectively).

In full agreement with the experimental and calculated results $D_{ls}$ for the $1^+$ ($T = 1$) state in $^{12}$C (see the Fig. 2) are the corresponding data for the same excitation at higher energies of 400–500 MeV (Fig. 1). Indeed, the experimental and calculated values $D_{ls}$ for the isovector excitation are still very small.

The isospin representation for the moduli of the spin-orbit components in the effective $NN$ interaction (parametrization of the free $NN$ $t$-matrix from Love and Franey) at $E_p = 140$ and 800 MeV is in qualitative agreement with the analyzed results. Undeniably, these isoscalar and isovector moduli, displayed as a function of $q$, have similar shapes. However, at all the $q$ quantities, the isoscalar (as against isovector) "interaction" components are typically dominant in value [5].

## 4. TRANSVERSE- AND LONGITUDINAL-SPIN-TRANSFER PROBABILITIES FOR UNNATURAL-PARITY $(\vec{p},\vec{p}')$ REACTIONS AT INTERMEDIATE ENERGIES

In this section, we present systematized angular distributions of two spin (polarization)-transfer probabilities $D_K$ (Fig. 3), based on the available complete $(\vec{p},\vec{p}')$ measurements at $E_p = 200$ MeV (solid points), as well as at $E_p = 350$ MeV, $E_p = 500$ MeV (open points for $^{16}$O and $^{28}$Si, respectively) for the unnatural-parity $T = 1$ levels in a set of light-weight nuclei. Instead of measured and calculated polarization-transfer coefficients, Fig. 3 shows their combinations, which as probabilities $D_K$, are associated in the PWIA prediction with the squares of particular amplitudes in the $NN$ effective interaction. The given $D_K$ are polarization observables introduced in [1].

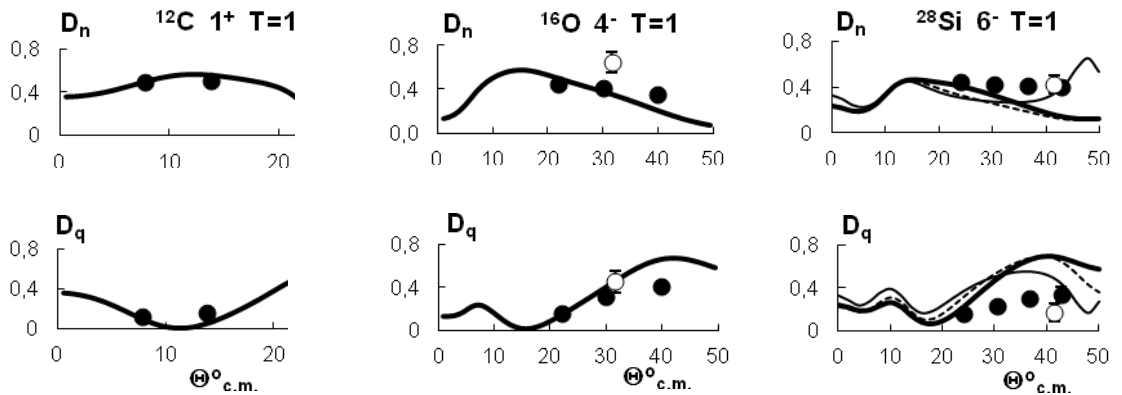

*Fig. 3. Systematized angular distributions of the spin-transfer probabilities, $D_n$ and $D_q$, for the indicated isovector levels in $^{12}$C, $^{16}$O and $^{28}$Si, based on the measured $(\vec{p},\vec{p}')$ quantities at 200 MeV [15, 16], as well as at 350 and 500 MeV. The measurements at 350 MeV ($^{16}$O) were performed in [7], and at 500 MeV ($^{28}$Si) – in [17]. The calculations applied to 200 MeV were as follows. In the case of $^{12}$C, we used DWIA calculations based on the DBHF interaction [15, 16]. For $^{16}$O, we employed the DWBA 91 program from Raynal and G-matrix (DD) from Geramb. For $^{28}$Si, all our calculations were made using the DWBA 91 program with three types of interactions: G-matrix from Geramb (thick solid curve) and two variations of the Idaho interaction (thin solid and dashed curves) [18].*

As it was noted above, a distinctive feature of the $T = 1$ excitations is that the isovector spin-orbit interaction is weak at intermediate energies. The smallest quantity of $D_{ls}$ both in experiments and in our calculations confirms this fact (Fig. 2). As a rule, normal spin-transverse $D_n$ quantities exceed $D_{ls}$ values (Fig. 3).

Consequently, the spin-observable combinations $D_n$ that depend on the isovector spin-spin interaction and the transverse spin-matrix element are generally well described by calculations for the excitations represented in Fig. 3: $1^+$ (15.11 MeV) in $^{12}$C, $4^-$ (18.98 MeV) in $^{16}$O, and $6^-$ (14.35 MeV) in $^{28}$Si.

The contrasting term $D_q$ is a spin-longitudinal component of the spin-transfer probability. The $D_q$ values depend on the spin-spin interaction, namely on the pion-dominated isovector $(\sigma_1 \cdot \hat{q}\, \sigma_2 \cdot \hat{q})$ piece of the $NN$ interaction, as well as on the longitudinal spin-matrix element.

Different $q$ dependences in $D_n$ and $D_q$ can be explained by the fact that the transverse and longitudinal axial form factors have a different $q$ dependence. The qualitative features of such a relative behavior of $D_n$ and $D_q$ can be understood from Fig. 4, which represents the isovector transverse ($V_T$) and longitudinal ($V_L$) parts of the $t$-matrix interaction at 210 MeV (see [19]). When $\theta_{cm}$ in Fig. 3 varies from 20° to 40°, this corresponds to $q$ changing from 1 to 2 fm$^{-1}$ (Fig. 4).

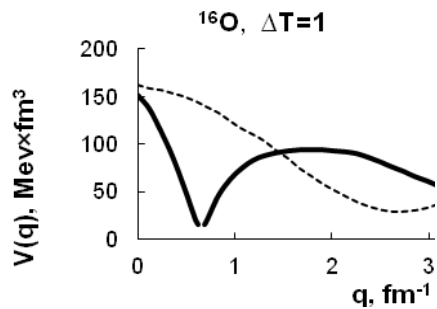

*Fig. 4. Transverse and longitudinal isovector parts of the t-matrix interaction at $E_p = 210$ MeV ($V_T$ and $V_L$, respectively) – from Love and Franey. $V_L$ is shown by a solid curve. $V_T$ is represented by a dashed curve.*

Let us apply these results to the $4^-$, $T = 1$ state in $^{16}$O. Thus, near 1.5 fm$^{-1}$, where $V_T \sim V_L$ (Fig. 4), $D_n$ and $D_q$ have similar quantities (Fig. 3), as the transverse and longitudinal transition densities are also comparable for the stretched $4^-$, $T = 1$ excitations [19]. Near 0.7 fm, where $V_L$ is very small, $D_q$ becomes roughly equal to 0. The $D_n$, however, acquires its maximum value, since here $V_T$ is large and dominant. At larger $q$, where $V_L$ becomes bigger than $V_T$, $D_q$ acquires larger values as compared to $D_n$. At 0° ($q = 0$), when $V_T$ and $V_L$ are practically the same (Fig. 4) we also observe similarity of $D_n$ and $D_q$ in our calculated data (Fig. 3).

Although isoscalar spin-dependent forces present a different picture, they can be analyzed in a similar manner. Let us note that the relation between the isoscalar $D_n$ and $D_q$ probabilities is entirely different for the $T = 0$ states (as compared to $T = 1$). We will show this relation in detail further on. Now we will only point out a unique sensitivity of the nucleon to the longitudinal spin response of the nucleus, which cannot be detected in the $e$- and $\pi$-nucleus interactions [1].

## 5. THE EXCHANGE PART OF UNNNATURAL-PARITY TRANSITIONS IN ($p,p'$) SCATTERING

As is shown above, the $(\vec{p}, \vec{p}')$ polarization-transfer measurements allow to establish the fact that the three spin-transfer probabilities $D_K$ are predominantly sensitive to the response of the nucleus through the tensor operators ($D_q$, $D_n$, and $D_p$), while the observable $D_{ls}$ is driven primarily by the spin-orbit operator.

As we observed in Fig. 1, the value and shape of the component $D_p$ can be reasonably predicted. At the same time, it is known that this characteristic is not simple. Indeed, the in-plane spin-transverse response $D_p$ is usually associated with tensor exchange amplitudes (see, e.g., [10]), therefore this specific quality of $D_p$ appears to be difficult to study.

One of the methods of such an analysis is based on a simultaneous description of the $(\vec{p}, \vec{p}')$ observable combinations and their ($e,e'$) counterparts as it is shown in [3]. In principle, the obtained information should allow to extricate from the $(\vec{p}, \vec{p}')$ data possible contributions of nuclei currents (convection and / or composite). The key element of such a study in a non-relativistic DWBA treatment (similar to the one we employed) is that the convection current arises exclusively through knock-on exchange and other nonlocalities [3, 13. 14].

A combined study of $(\vec{p}, \vec{p}')$ and $(e,e')$, using both relativistic and nonrelativistic analyses [3], showed the $(\vec{p}, \vec{p}')$ quantities can be relatively insensitive to the knock-on exchange. Therefore, the convection-composite currents are negligible in the $1^+$, $T = 1$ (15.11 MeV) excitation in $^{12}C$.

However, as is noted in [3], for the $1^+$, $T = 0$ (12.71 MeV) isoscalar transition in $^{12}C$, the analysis indicates that the convection-composite-current contributions are relatively much more important than they are for the $1^+$, $T = 1$ $(\vec{p}, \vec{p}')$ isovector transition. Based on this characteristic, especially prominent are the measured $T = 0$, $(\vec{p}, \vec{p}')$ quantities, related to the response $I_0 D_p$ and $I_0 D_{ls}$, where the factor $I_0$ represents an ordinary unpolarized cross section.

In a number of our works in order to evaluate direct and exchange pieces of the effective interaction, we compared the results of the analysis of $(\vec{p}, p')$ or $(\vec{p}, \vec{p}')$ spin observables that were obtained with the use of two calculation models. One of these models, represented in the LEA program from Kelly, is based on the zero-range treatment of knock-exchange. To oppose this approach, we employed another, a more complex model that involves the use of a finite-range treatment for the exchange part of the scattering. In this case, it was necessary to solve a more serious problem of establishing the exact nucleon-nucleus kinematics.

To perform such exact calculations, we used the program DWBA 91 from Raynal, and in some cases, we checked them against our calculations based on the data from Stephenson. In the latter case, the DWBA 86 code was used, as a program with finite-range DWIA (based on the works of Schaeffer and Raynal).

Having compared the experimental data and the results of the analysis with the zero-range calculations of LEA and those performed with the finite-range DWBA 91 program, we successfully obtained two sets of polarization transfer coefficients $D_{ij}$ for $(\vec{p}, \vec{p}')$ reactions leading to unnatural-parity $1^+$, $T = 0$ and $T = 1$ states in $^{12}C$ [20].

It follows from the analysis of the combinations $D_K$ for the $1^+$, $T = 0$ and $T = 1$ states in $^{12}C$, using both programs (DWBA 91 and LEA), that for these purposes the approach that do not represent the full $NN$ amplitude cannot be used in certain cases. This applies to the program LEA in which the exchange approximation is such that the exchange interactions are independent of the momentum transfer and are reduced to delta functions [21].

The most striking discrepancy between the experimental and calculated data within the framework of LEA formalism is established by us in the case of the $1^+$, $T = 0$ excited state in $^{12}C$ for the combination $D_p$ (Fig. 5, dashed line) and $D_{ls}$ (not shown) at $E_p = 200$ MeV.

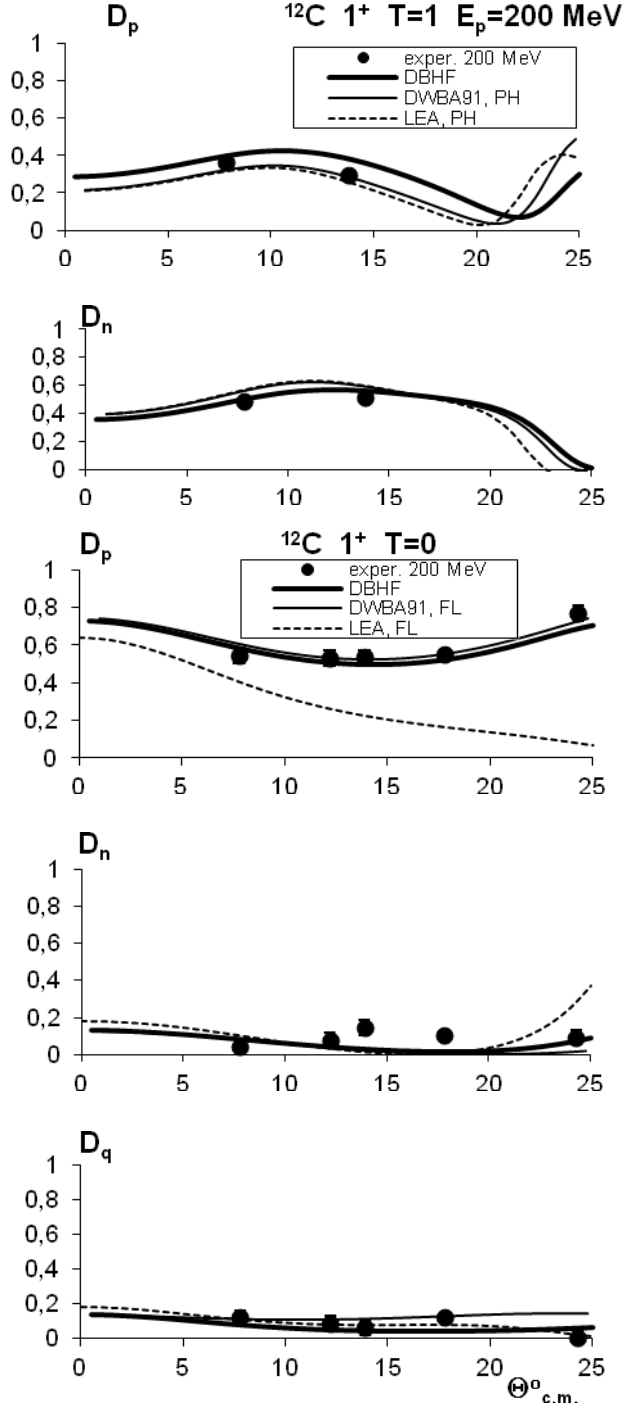

***Fig. 5.*** *Sensitivity of $D_K$ to the direct contributions and the expected role of nonlocal processes. All the solid lines (both thick and thin) represent the calculations with the exact finite-range amplitudes (two versions of the DWBA programs from Raynal); the dashed curves are the zero-range calculations of LEA. The lower part represents the observables for the $1^+$, $T = 0$ state, and the upper part represents the observables for the $1^+$, $T = 1$ state in $^{12}C$. The thick solid curves are DWBA 86 calculations using the DBHF interaction of Sammarruca and Stephenson, while the thin solid curves use DWBA 91 calculations with the free interaction of Franey and Love ($T = 0$) and the effective PH interaction of Geramb and his Hamburg associates, based upon the Paris potential ($T = 1$). This potential is used in all LEA calculations. The combination of experimental data is based on the measurements from [15].*

However, there is no such discrepancy between similar calculated and experimental data in the case of $1^+$, $T = 1$ excited state in the same nucleus, $^{12}$C, for the combination $D_p$ (Fig. 5, dashed line) and $D_{ls}$ (not shown) at the same $E_p$ value.

All the results of the analysis are in good agreement with the above-mentioned studies [3] aimed at extracting information on $M1$ spin responses from $(e,e')$ measurements and the analogous $(\vec{p},\vec{p}')$ quantities at $E_p = 500$ MeV, respectively. Based on the combined analysis of the indicated data, the conclusion about the corresponding convection-current contributions, and, therefore, about the role of knock-on exchange was made in [3]. It is worth noting that in the $I_0D_p$ and $I_0D_{ls}$ data for the $1^+$, $T = 0$ state in the $^{12}$C excitation, the indicated contributions turned out to be relatively more important than they are for the $1^+$, $T = 1$ transition. This, in particular, is in agreement with our analysis discussed above and represented in Fig. 5.

Moreover, the research results, systematized in [22], clearly indicate that for the inelastic excitation of the $1^+$, $T = 0$ state in the $^{12}$C by inelastic proton scattering, the tensor exchange process is a dominant reaction mechanism, while the direct process is suppressed. In this connection it is clear that the DWBA 91 calculations describe $D_p$ for the $1^+$, $T = 0$ state (Fig. 5, solid curves) and $D_{ls}$ for the $1^+$, $T = 0$ state (not shown) fairly well. Moreover, antisymmetrization introduces additional nonlocalities in the form of knock-on exchange amplitudes, as it occurs in the case of nucleons [5].

Nevertheless, it is also seen (Fig. 5) that for a number of the observable combinations $D_K$ the zero-range treatment of knock-on exchange in LEA is not inferior, as regards to the quality of the description of experiments, towards the exact finite-range calculations made with the use of the DWBA 91 or DWBA 86 programs.

To ensure a better understanding of the analytical picture represented in Fig. 5, let us consider the previous results of the analysis of the $(\vec{p},\vec{p}')$ polarization transfer measurements (the polarized cross sections) for the $6^-$, $T =1$ transition in $^{28}$Si $(\vec{p},\vec{p}')$$^{28}$Si [10]. There experimental data are compared with two types of calculations. One of them is based on the original DWIA program LEA [21], the same one as we used (Fig. 5). The other type of calculations is based on a modification that agrees well with the exact finite-range treatment, i.e., basically, the same that we employed for two types of analysis represented in Fig. 5, using the DWBA 91 and DWIA 86 programs. If we consider the analytical data, represented in Fig. 5, from the angle of those studies, we will observe the following similarity. Thus, the spin-transfer response $\sigma_p \equiv I_0D_p$ for the $6^-$, $T =1$ transition in $^{28}$Si [10], and spin-transverse component $D_p$ for $1^+$, $T = 1$ excited state in $^{12}$C (Fig. 5) are in good agreement with the calculations of both types indicated above.

Besides, it should be noted that there also exist relatively small differences between the data and calculations using both $NN$ matrixes,

considered above, in the case of the response $I_0D_n$ [10] and the component $D_n$ (Fig. 5) for the same isovector $(\Delta T = 1)$ transitions, respectively.

However, in the case of the $I_0D_q$ response and the component $D_q$ for the same transitions and in the same nuclei, the situation is very different. Therefore, despite the fact that the $D_q$ data still agree equally well with the calculations of both types, for the $I_0Dq$ response (in the case of $6^-$, $T=1$ in $^{28}$Si) only finite-range calculations are necessary [10]. Besides, our analysis of the spin-transfer responses $D_n$ and $D_q$ for the $6^-$, $T=1$ state in $^{28}$Si shows that the calculated results (Fig. 3) underestimate $D_n$ and overestimate experimental data for $D_q$, which largely corresponds to the estimations from [16].

As we noted above, in a nonrelativistic framework unnatural-parity transitions appear to be sensitive to the coupling of the nucleon spin to the bound current, with these couplings arising through exchange processes.

It is known now that the $(p,p')$, $(\vec{p}, p')$ and $(\vec{p}, \vec{p}')$ observables and some of their combinations are sensitive, to a different extent, to the nonlocal and exchange nature of the scattering process. Moreover, there may be several sources of nonlocality in the spin-dependent parts of the effective $NN$ interaction.

According to [14], although a quantitative understanding of the sources of nonlocality in the spin-dependent parts of the $NN$ interaction is rather difficult, it is important to find a signature of their presence. Such examples as those represented in Fig. 5 help to discriminate between possible $NN$ effective interactions with different nonlocal behavior.

## SAMMARY AND CONCLUSIONS

As we have established it for the spin-transfer response $D_K$, part of which is shown in Fig. 5, in some cases exchange processes become especially important for transitions with $\Delta T = 0$ in the $^{12}$C nucleus. At the same time, for other quantities (particularly for transitions with $\Delta T = 1$ in the same nucleus) the role of exchange processes is much weaker, judging by the fact that there the zero-range treatment of knock-on exchange in LEA works as satisfactorily as exact finite-range calculations in DWBA 91.

Such a result for the $\Delta T = 1$ transition in $^{12}$C is in good agreement with the conclusions of work [3] in the case of 500 MeV. The authors of this paper showed that in the case of $\Delta T = 1$ the $(\vec{p}, \vec{p}')$ quantities were relatively insensitive to the knock-on exchange, and the convection-composite currents were confirmed to be very small.

We conducted our analysis at noticeably smaller energies ($E_p = 200$ MeV) because with such energies exchange (nonlocal) processes are expected to be more significant in strength as the incident proton energy decreases (see specifically [23]).

In fact such expectations proved to be true, although most clearly only for the $\Delta T = 0$ transition. Here our nonrelativistic calculations with exchange showed that in some cases experiments could be carried out only with *t* (or *G*) matrix used to represent the full *NN* amplitude. In any case we did not get anywhere near the "curious situations" at any $\Delta T$ values as those observed in [23] where relativistic codes were used, so, to achieve a better description of the data, the authors had to exclude knock-on exchange contributions altogether.

The main approach we used in this work was that in the framework of the model represented in [1], according to [5, 16], the complete $(\vec{p}, \vec{p}')$ spin-transfer observables can be used for study to isolate specific parts of the nuclear response. This is supported by the fact that the combinations of polarization-transfer coefficients selectively (separately) depend on individual (particular) components of the *NN* amplitude.

Criticism regarding this method [24] primarily concerns the point that certain combinations of $(\vec{p}, \vec{p}')$ spin-transfer observables and their relation to individual terms in the *NN* interaction are linked to a direct plane-wave impulse approximation. Therefore, according to [24], the inclusion of distortion of projectile waves and knock-on exchange blurs the physical interpretation of corresponding nuclear response functions.

However, as was shown above, such doubts can largely be overcome. Thus, as it is suggested in [5], this method allows to obtain serious information about the transverse and longitudinal spin-response functions of the nucleus. Indeed, the comparison of the corresponding quantities $D_n$ and $D_q$ for the states $1^+$, $T = 0$ (Fig.5) and $1^+$, $T = 1$ (Figs. 1 and 3) demonstrates their obvious isospin separation.

Good agreement between two DWBA calculations shown in Fig. 5 by thick and thin lines, respectively, should also be pointed out. The thin solid lines were obtained within several variations of the approach that formulated the proton-nucleus scattering problem in the nonrelativistic impulse approximation. The second approach (thick solid lines) represents the DBHF technique based on the Dirac–Brueckner theory. There the authors followed closely the relativistic approach to nuclear matter [16]. As is seen from Fig. 5, there is no significant discrepancy between these two types of calculations.

In [8] we described a number of difficulties in coordinating calculations and experiments for the transverse and longitudinal responses of the stretched states $4^-$, $T = 0$ in $^{16}O$ and $6^-$, $T = 0$ in $^{28}Si$. It is evident that the theoretical picture presented there was too simplified in comparison with the experiment.

Nevertheless, at the excitation of states with the same spins and parities in the same nuclei, but for the $\Delta T = 1$ transitions, such responses can be described rather adequately (Fig. 3). Besides, well known are the facts that for the unnatural-parity transition, the spin-orbit part of the effective interaction is small for isovector transitions, and the spin-orbit amplitude is large in the isoscalar channel, which is effectively confirmed in Fig. 2. Therefore, we quite agree with the assertion of the authors of work [16] that such representations can be used as a powerful diagnostic tool. However, many unsolved problems still remain, and a more comprehensive investigation is needed in the study of polarization-transfer processes in $(\vec{p}, \vec{p}')$ reactions.